\documentclass[showpacs,showkeys,preprint]{revtex4}
\usepackage{graphicx}

\def\ii{\'\i}

\begin{document}

\title{Validity of the Brunet-Derrida formula for the speed of pulled fronts with a cutoff}
\author{R. D. Benguria, M. C. Depassier}
\affiliation{
 Facultad de F\'\i sica\\
     Pontificia Universidad Cat\'olica de Chile\\
           Casilla 306, Santiago 22, Chile}
\author{M. Loss}
\affiliation{
 School of Mathematics, Georgia Tech,
Atlanta, GA 30332}

\date{\today}

\begin{abstract}
We establish rigorous upper and lower bounds for the speed of pulled fronts with a cutoff. We show that the Brunet-Derrida formula corresponds to the leading order expansion in the cut-off parameter of both the upper and lower bounds. For sufficiently large cut-off parameter the Brunet-Derrida formula lies outside the allowed band determined from the bounds.  If nonlinearities are neglected the upper and lower bounds coincide and are  the exact linear speed for all values of the cut-off parameter.
\end{abstract}


\maketitle

The reaction diffusion equation
\begin{equation}
u_t = u_{xx} + f(u)
\label{rd}
\end{equation}
provides a simple description of  phenomena in  fields  such as  population dynamics, chemical reactions, flame propagation, fluids, QCD, among others \cite{Murray,Britton,Showalter,Zeldovich,Munier}. It is one of the simplest models which shows  how a small perturbation to an unstable state develops into a moving front joining a stable to an unstable state. The reaction term $f(u)$ satisfies different conditions depending on the physical problem of interest. One of the first, and most studied cases, is the Fisher reaction term
$f(u) = u (1 -u)$ for which  the asymptotic speed of the propagating
front is $c=2$, a value determined from linear considerations.
 A more general case was studied by Kolmogorov, Petrovskii and Piscounov (KPP)\cite{KPP} who showed that for all reaction terms which satisfy the KPP condition
\begin{equation}
f(u) > 0, \qquad f(0) = f(1) = 0, \qquad  f(u) < f'(0) u
\label{KPPcondition}
\end{equation}
the asymptotic speed of the front joining the stable $u=1$ point  to the unstable $u=0$ point is given by
$$
c_{KPP} = 2 \sqrt{f'(0)}.
$$
These fronts are called pulled since it is the leading edge of the front which determines the velocity of propagation. In the rest of this work we assume that $f'(0)=1$.
The evolution of localized initial conditions for general reaction terms, and rigorous  properties of the fronts were studied by  Aronson and Weinberger \cite{AW78}. The  asymptotic speed of the front for all reaction terms can be found from the integral variational principle \cite{BD96c}
\begin{equation}
c^2 = \sup_{g(u)} \,2 \,\frac
{ \int_0^1 f(u) g(u) d\,u}
{\int_0^1 g^2(u)/h(u) d\,u}
\label{vp1}
 \end{equation}
where the supremum is taken over all positive monotonic decreasing functions $g(u)$ for which the integrals exist and where $h(u) = -g'(u)$. The supremum is always attained for reaction terms which are not pulled.

Two effects not included in the classical reaction diffusion equation (\ref{rd}), are the effect of noise and the effect of a finite number $N$ of diffusive particles. It was shown by Brunet and Derrida that such effects can be simulated by introducing a cut-off in the reaction term. In the case of noise the cut-off parameter measures the amplitude of the noise while in the case of finite number of $N$ diffusing particles the cut-off parameter $\epsilon = 1/N$. There is substantial numerical evidence that introducing a cut-off in the reaction terms reproduces accurately the effect of noise and finiteness in the number of diffusing particles \cite{brunet-derrida}.

By means of an asymptotic matching Brunet and Derrida  showed that for a reaction term $f(u) = u(1 - u^2)$ a small cut-off changes the speed of the front to
\begin{equation}
c \approx 2 -\frac{\pi^2}{(\ln \epsilon)^2}.
\label{brunet}
\end{equation}
In recent work it has been show that the Brunet-Derrida formula for the speed is correct to $\cal{O}$$((\ln\epsilon)^{-3})$ for a wider class of reaction terms \cite{Kaper}.

The purpose of this work is to show that for reaction terms  of the form $f(u) \Theta(u-\epsilon)$ where $f$ satisfies the KPP condition Eq. (\ref{KPPcondition}) and $\Theta$ is the step function,
the speed $c$ of the front with the cutoff satisfies
\begin{equation}
2 \sin (\phi_*) - \Delta(\phi_*) < c < 2 \sin ( \phi_*),
\label{main}
\end{equation}
with
\begin{equation}
\phi_* \, \tan( \phi_*) = \frac{1}{2} \vert \ln(\epsilon) \vert
\end{equation}

 We see that for $0<\epsilon< 1$, $0< \phi_* < \pi/2$. The function  $\Delta(\phi_*)$ depends on the nonlinear terms of the reaction function. For small $\epsilon$ the series expansion of the upper bound $c_{UP}$ is
\begin{equation}
c_{UP} = 2 \sin(\phi_*) = 2 -\frac{\pi^2}{(\ln \epsilon)^2} + {\cal{O}} ((\ln\epsilon)^{-3})
\end{equation}
The contribution of the nonlinearities, contained in the term $\Delta(\phi_*)$, appears at $\cal{O}$$((\ln\epsilon)^{-3})$, so that the leading order terms in the expansion  of the upper and lower bounds give  the Brunet-Derrida formula. If nonlinearities are neglected the value $2 \sin ( \phi_*)$ is the analog of the KPP value $c=2$ for reaction terms which satisfy the KPP condition, but with a cutoff.

In what follows we derive the bounds and apply them to the  Fisher reaction term \cite{Fisher} {$f(u) = u-u^2$} and to the reaction term studied by Brunet and Derrida $f(u) = u - u^3$.
The main tool to obtain the bounds is the variational principle for the speed.

As shown in previous work \cite{BDPRE07}, we may perform the  change variables  $
u = u(s)$ where  $s= 1/g$ in Eq.(\ref{vp1}) and write the variational expression for the speed as
\begin{equation}
c^2 = \sup_{u(s)} 2\, \frac{ F(1)/s_0 + \int_0^{s_0} F(u(s))/s^2 d\,s}{\int_0^{s_0} \left( d u /d s\right)^2 d\,s},
\label{newvp}
\end{equation}
where $s_0= 1/g(u=1)$ is an arbitrary parameter,
$$
F(u) = \int_0^u f(q) dq.
$$
and  the supremum is taken over positive increasing functions $u(s)$ such that $u(0)=0$, $u(s_0) =1$ and for which all the integrals in (\ref{newvp}) are finite. Therefore, for any suitable trial function $u(s)$ we know that

\begin{equation}
c^2 \ge 2\, \frac{ F(1)/s_0 + \int_0^{s_0} F(u(s))/s^2 d\,s}{\int_0^{s_0} \left( d u /d s\right)^2 d\,s}.
\label{lowerbound}
\end{equation}

Consider now reaction terms $f(u)$ with a cut-off $\epsilon$ of the form
$$
f(u) = \left\{ \begin{array}{ll}
              0
             & \mbox{if $0\leq u \leq \epsilon$} \\
             u - N(u) &\mbox{if $\epsilon < u < 1,$}
           \end{array} \right.
$$
where $N(u)$, the nonlinearity, is such that $N(0)= N'(0) =0$. We find
$$
F(u) = \left\{ \begin{array}{ll}
              0
             & \mbox{if $0\leq u \leq \epsilon$} \\
             u^2/2 -  \epsilon^2/2 + F_n(u) &\mbox{if $\epsilon < u < 1$},
           \end{array} \right.
$$
where $F_n(u) = - \int_\epsilon^u N(u) d u.
$

Assume now that $f(u)$ satisfies the KPP criterion Eq.(\ref{KPPcondition}). Since $f(u) < u$, it follows that $F(u) \le G(u)$ where
$$
G(u) = \left\{ \begin{array}{ll}
              0
             & \mbox{if $0\leq u \leq \epsilon$} \\
             u^2/2 -  \epsilon^2/2  &\mbox{if $\epsilon < u < 1$},
           \end{array} \right.
$$
and therefore
\begin{equation}
c^2  <  {\cal{G}}[u]  \equiv  \sup_{u(s)} 2 \,\frac{ G(1)/s_0 + \int_0^{s_0} G(u(s))/s^2 d\,s}{\int_0^{s_0} \left( d u /d s\right)^2 d\,s}.
\end{equation}
One can prove (rigorous details will be given elsewhere) that  ${\cal {G}}$ is bounded above and that there exists a function $\hat u(s)$ for which the supremum is attained. This function is the monotonic increasing solution to the Euler-Lagrange equation for
${\cal {G}}$ satisfying the boundary conditions $\hat u(0) =0, \hat u(s_0) =1$. One can also prove that the variational parameter $s_0$ is finite and $\hat u'(s_0) = 0$. In summary, the maximizing function for
${\cal {G}}$ is the solution of
$$
\frac{d^2 \hat u}{d s^2} =0  \qquad \qquad {\rm for} \qquad 0< \hat u< \epsilon.
$$
$$
 \frac{d^2 \hat u}{d s^2} + \lambda \frac{\hat u} {s^2} =0,\qquad \qquad {\rm for} \qquad \epsilon < \hat u< 1
 $$
 subject to the boundary conditions
 $$
 \hat u(0) =0, \qquad \hat u(s_0) = 1 \qquad  \hat u'(s_0)=0 \qquad \hat u'(s) >0,
 $$
with the function and its derivative continuous at $\hat u = \epsilon$.

The solution to this problem is given by
$$
\hat u(s) = \left\{
\begin{array}{ll}
              s
             & \mbox{if $0\leq s \leq \epsilon$} \\
           A \sqrt{s} \cos(\phi(s)) &\mbox{if $\epsilon < s< s_0$},
           \end{array} \right.
$$
with
\begin{equation}
 A =\sqrt{\epsilon} \sec (\phi_*),\qquad \qquad s_0 = 1/\epsilon, \qquad \qquad  \phi(s) = \frac{1}{2} \cot (\phi_*)\ln (s/\epsilon) - \phi_*
\end{equation}
 where $\phi_*$ is the first positive solution of
\begin{equation}
\phi_* \tan \phi_* = \frac{1}{2} |\ln \epsilon|.
\label{valor}
\end{equation}

The maximum of ${\cal{G}} ={\cal{G}}[\hat u]$ can be calculated easily. We obtain after performing the integrals,
\begin{equation}
c^2  <  {\cal{G}}[\hat u] = 4 \sin^2 (\phi_*) \equiv c^2_{UP}.
\label{top}
\end{equation}

To obtain the lower bound we shall use the optimizing function $\hat u(s)$ as a a suitable trial function in Eq. (\ref{lowerbound}). We obtain
\begin{equation}
c^2 \ge 4 \sin^2 (\phi_*) + \frac{4 \sin(\phi_*) \cos^3(\phi_*)}{\epsilon( 2 \phi_* + \sin(2 \phi_*))} \left[ \epsilon F_n(1) + \int_\epsilon^{1/\epsilon} F_n(\hat u(s))/s^2 d\,s\right]
\label{bottom}
\end{equation}

Since $F_n$ is negative, we may combine Eqs.(\ref{top}) and (\ref{bottom}) and write our main result as given in Eq. (\ref{main}).

As an example consider the reaction term studied by Brunet and Derrida, $f(u) = u - u^3$. The lower bound can be written explicitly as
$$
c^2 > 4 \sin^2(\phi_*) - \frac{2 (1 - \epsilon^2)\cos^3(\phi_*)\sin(\phi_*)}{2 \phi_* + \sin(2 \phi_*)} -
\frac{2 \epsilon \sin(\phi_*)}{\cos(\phi_*) (2 \phi_* + \sin(2 \phi_*))} \int_\epsilon^{1/\epsilon} \cos^4 \phi(s) d\,s.
$$
The integral has a long analytic expression which we omit here. From the explicit expression above it is not difficult to show that the contribution of the two last terms, which arise from the nonlinear terms, are of ${\cal{O}}(|\ln\epsilon|^{-3}|$.
In figures 1 and 2 we show the bounds together with the Brunet-Derrida formula as a function of $\epsilon$. The solid lines correspond to  the upper and lower bounds. The dashed line is the Brunet-Derrida formula.
\begin{figure}[h]
    \includegraphics[height=5.0cm]{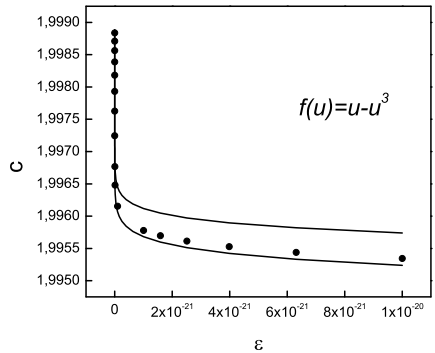}
    \caption{Speed as a function of the cut-off parameter for the reaction term $f(u)=u-u^3$. The solid lines correspond to the bounds, the dots to the Brunet-Derrida formula. }
\end{figure}

\begin{figure}[h]
    \includegraphics[height=5.0cm]{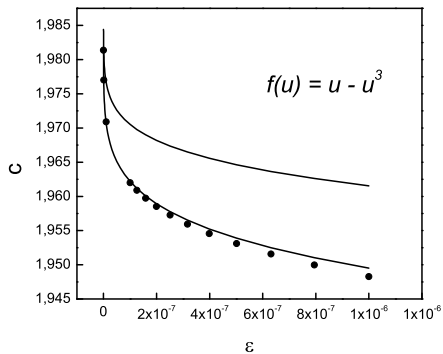}
    \caption{As in Fig. 1 for different values of the cut-off parameter. }
\end{figure}

As a second example we consider the Fisher reaction term $f(u) = u (1-u)$ with a cut-off. The lower bound becomes
$$
c^2 >   4 \sin^2(\phi_*)- \frac{8}{3}\frac{(1 - \epsilon) \sin(\phi_*) \cos^3(\phi_*)}{(2 \phi_* + \sin(2 \phi_*))} -
     \frac{8}{3}\frac{\sqrt\epsilon \sin(\phi_*)}{(2 \phi_* + \sin(2 \phi_*))} \int_\epsilon^{1/\epsilon} \frac{\cos^3 \phi(s)}{\sqrt{s}} d\,s.
$$
Again, the integral can be done analytically and we do not show it here.

In Fig. 3 we show the upper and lower bounds and the Brunet-Derrida formula. In this case the Brunet-Derrida formula leaves the allowed band at larger value of $\epsilon$. In general for reaction terms $f(u) = u - u^n$, the gap between the upper and lower bounds becomes narrower and the Brunet-Derrida formula valid for a smaller range of $\epsilon$.

\newpage
\begin{figure}[h]
    \includegraphics[height=5.0cm]{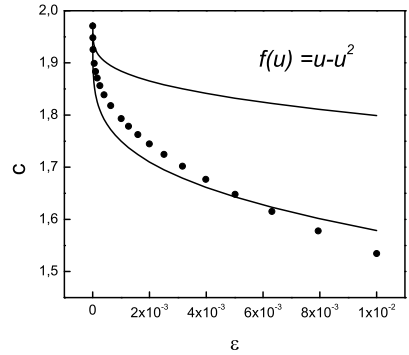}
    \caption{Speed as a function of the cut-off parameter for the reaction term $f(u)=u-u^2$. Lines as in Fig. 1 }
\end{figure}

{In summary, we have studied the effect of a cut-off on reaction terms which satisfy the KPP condition Eq.(\ref{KPPcondition}). We have found upper and lower bounds valid for all values of the cut-off parameters, which allow to assess the accuracy of the Brunet Derrida formula. If we consider only the linear terms, the upper and lower bounds coincide and give the exact linear value for the speed, of which the two leading order terms are the Brunet-Derrida formula.}

\section*{Acknowledgements}
We acknowledge partial support of Fondecyt (CHILE) projects 106--0627 and 106--0651,  CONICYT/PBCT Proyecto Anillo de Investigaci\'on en Ciencia y Tecnolog\ii a ACT30/2006 and
U.S. National Science Foundation grant DMS 06-00037.

\end{document}